\documentclass[fleqn,twoside]{article}
\usepackage{espcrc2,epsfig}

\usepackage{graphicx}
\usepackage[figuresright]{rotating}

\newcommand{\AmS}{{\protect\the\textfont2
  A\kern-.1667em\lower.5ex\hbox{M}\kern-.125emS}}

\title{Precision Electroweak Measurements Circa 2002
\thanks{Talk given at the ICHEP Conference, Amsterdam
                                               24-31 July
2002. Preprint  {\sf CERN-TH/2002-311}}}

\author{Paolo Gambino\address[MCSD]{
Theory Division - CERN\\
CH 1211, Geneve 23, Switzerland}
        \thanks{Marie Curie Fellow}}
\begin{document}

\begin{abstract}           
Present global fits to electroweak data are characterized by two
results that differ from Standard Model (SM) expectations by about $
3 \sigma$, the NuTeV measurement of $\sin^2\theta_W$  and the FB
$b$ quark asymmetries measured at
LEP. I review possible SM and new physics explanations of these
anomalies and the  implications  for the indirect
determination of the Higgs mass.
\vspace{1pc}
\end{abstract}

\maketitle

\section{The global Standard Model  fit}
\newcommand{\smallw}{{\scriptscriptstyle W}} %
\newcommand{\smallh}{{\scriptscriptstyle H}} 
\newcommand{\smallp}{{\scriptscriptstyle P}} %
\newcommand{\mh}{M_\smallh}
\newcommand{\mw}{M_\smallw}
\newcommand{\mt}{M_t}
\newcommand\gsim{\mathop{\mbox{\vbox{\hbox{$>$} \vskip -9pt \hbox{$\sim$}
             \vskip -3pt  }}}}
\newcommand\lsim{\mathop{\mbox{\vbox{\hbox{$<$} \vskip -9pt \hbox{$\sim$}
             \vskip -3pt  }}}}

There is not so much going on in electroweak physics nowadays, apart
from the muon $g-2$, it is
tempting to say. Not quite so:
the latest Standard Model (SM) fit performed by the LEP 
Electroweak Working Group \cite{martin} looks remarkably different
from the one from last year. The main new result comes from the NuTeV
Collaboration \cite{nutev}:
their measurement of the electroweak  mixing angle in $\nu$-N Deep
Inelastic Scattering (DIS) 
differs about 3$\sigma$ from theoretical expectations.
The  $\chi^2/$d.o.f.\ of the global fit is 29.7/15, corresponding to 
1.3\% probability. The NuTeV result shares the responsibility for the
degradation of the fit  
with another deviant measurement, that of the bottom quark
Forward-Backward asymmetry, $A_{FB}^b$, at LEP. The best fit
\cite{martin} points to a fairly
light Higgs boson, with mass $\mh=81$~GeV, while the 95\% CL upper
bound on $\mh$, including an estimate of theoretical uncertainty,
 is about 190~GeV.
Interestingly, the information on the Higgs mass is almost insensitive
to the NuTeV result: a fit performed excluding this new result gives
practically the same constraints on $\mh$, but of course the quality
of the fit improves significantly,  with $\chi^2/$d.o.f.=20.5/14,
corresponding to a  probability of 11.4\%. 
One would  conclude that the SM fit is quite satisfactory, if
not for NuTeV. Let us therefore start this (incomplete) status review
with a look at the intriguing NuTeV  {\it anomaly}.

 \section{The  NuTeV electroweak result}
 NuTeV measures ratios of Neutral  (NC) to Charged Current (CC)
 cross sections in $\nu N$
 DIS. Ideally, in the parton model with only one generation of quarks
and an isoscalar target
\begin{eqnarray}
&&R_\nu \equiv \frac{\sigma(\nu { N}\to \nu X)}{\sigma(\nu { N}\to \mu X)} =
g_L^2 + { r} g_R^2\nonumber\\
&&R_{\bar{\nu}} \equiv \frac{\sigma(\bar\nu { N}\to \bar\nu
X)}{\sigma(\bar\nu { N}\to \bar\mu X)} =
 g_L^2 + \frac{1}{ r} g_R^2,\label{rdef}
\nonumber
\end{eqnarray}
where
$ r \equiv  \frac{\sigma(\bar{\nu}{ N}\to \bar\mu
X)}{\sigma({\nu}{ N}\to \mu X)}$ and $g_{L,R}^2$ are average
effective left and right-handed $\nu$-quark couplings. 
The actual experimental  ratios
$R_{\nu,\bar{\nu}}^{exp}$ differ from $R_{\nu,\bar\nu}$ because
of $\nu_e$  contamination, experimental cuts, NC/CC misidentification,
the presence of second generation
quarks, the non-isoscalarity of steel  target, QCD and electroweak
corrections {\it etc.} In the NuTeV analysis, a
MonteCarlo including most of these effects 
relates $R_{\nu,\bar{\nu}}^{exp}$ to $R_{\nu,\bar{\nu}}$. 
It is useful to note that 
most uncertainties and $O(\alpha_s)$ effects drop in the
Paschos-Wolfenstein (PW) ratio \cite{PW}
\[\label{eq:PW}
R_{\smallp\smallw}\! \equiv\!\frac{R_\nu - { r} R_{\bar{\nu}}}{1- r}\! =\!
\frac{\sigma(\nu { N}\to \nu X)-\sigma(\bar\nu { N}\to
\bar\nu X)}{\sigma(\nu { N}\to \ell X) - 
\sigma(\bar{\nu}{ N}\to \bar{\ell}X)}
\]
which equals ${ g_L^2- g_R^2} %
= \frac{1}{2}-\sin^2 \theta_{\rm W}$ and therefore could  provide a clean
measurement of $\sin^2 \theta_{\rm W}$, if experimentally accessible.
NuTeV do not measure $R_{\smallp\smallw}$ directly,
but, using the fact that
$R_{\bar{\nu}}$ is almost insensitive to $\sin^2 \theta_{\rm
W}$, they  extract from it 
the main hadronic uncertainty, an effective charm mass. 
The weak mixing angle 
is then obtained from 
$R_{\nu}$.
In practice, NuTeV fit for $m_c^{eff}$ and $\sin^2 \theta_{\rm W}$.
 To first approximation, the NuTeV procedure
corresponds to a measurement of $R_{\smallp\smallw}$.
The result is expressed as a test on the on-shell 
$s_W^2\equiv 1- M_W^2/M_Z^2$ definition of $\sin^2\theta_W$:
\begin{equation}\label{result}
s_W^2({\rm NuTeV})=0.2277\pm0.0013\pm0.0006\pm0.0006,
\end{equation}
where the three errors are 
statistical, systematic, and theoretical, respectively.
Because of accidental cancellations, the choice of the on-shell scheme 
implies very small top and Higgs mass dependence in the above equation.
The above value must be compared to the one obtained 
using the results 
of the global fit, 
 $s_W^2=0.2226\pm 0.0004$, which is about 3$\sigma$ away. 

QED corrections are important and their implementation in NuTeV 
could 
be improved, but they seem at the moment an unlikely  explanation.
Electroweak corrections, on the other hand, are small and under control.

A potentially very important source of  uncertainty 
are the parton distribution functions (PDFs) employed in the analysis.
NuTeV work at Leading Order (LO) in QCD in the context of a {\it cross
  section model} which effectively introduces
 some Next to Leading Order (NLO) improvement. They use
LO PDFs self-consistently fitted in the experiment, with little 
external input. 

Is the NuTeV estimate of the PDFs uncertainty reliable?
We have seen  that  $R_{\smallp\smallw}$ is independent of the
details of first generation PDFs. 
As long as the NuTeV result is
equivalent to a measurement of $R_{\smallp\smallw}$, even with cuts
and second generation quarks, the small uncertainty
attributed by NuTeV might be  realistic \cite{noi}. The  problem is 
that NuTeV do {\it not} really measure  $R_{\smallp\smallw}$ and
there are indications \cite{noi} that this might be relevant at the required 
level of accuracy. 

We have seen that NuTeV do not employ NLO QCD corrections. Are they
necessary? The answer is very similar to the previous one: no, if you are 
measuring $R_{\smallp\smallw}$, which is not corrected at $O(\alpha_s)$.
But any CC/NC or $\nu/\bar{\nu}$ asymmetry (introduced by cuts, 
differences in the energy spectra and in the  sensitivity, {\it etc.}) spoils
delicate cancellations (ordinary NLO corrections are 5-10\%, while
here a better than 0.5\% precision is required).  
As the 
NuTeV measurement seems to differ enough from that of $R_{\smallp\smallw}$,
the analysis needs to be consistently upgraded to NLO. 
This would allow the implementation of
different sets of NLO PDFs, and would  simplify the discussion
of other issues, such as the PDFs uncertainty and the
contribution of an asymmetric quark sea.

{\subsection{ Asymmetric sea}}
In the previous section I have implicitly used the assumptions, 
generally made in the extraction of PDFs from the data,  
of isospin symmetry and of a symmetric strange and charm sea ($s=\bar
s$, $c=\bar c$).
If we drop these assumptions, the PW relation is explicitly violated by new
 terms \cite{noi}
\begin{equation}
R_{\smallp\smallw}= 
\frac12 -s_W^2 + \frac{\tilde{g}^2}{Q^-} \,( u^- - d^- +c^- - s^-),\label{viol}
\end{equation}
\vspace{-1mm}
where $q^-$ is 
the asymmetry in the momentum carried by the quark species $q$ 
in an isoscalar target, $q^-=\int_0^1  x \,[q(x)- \bar{q}(x)] \,dx$,
$\tilde{g}^2\approx .23$ a coupling factor, 
and $Q^-=(u^-+d^-)/2\approx 0.18$.
While there is no reason in QCD to expect $s^-\!=\! 0$, 
 for an isoscalar target $u^-\!-\!d^-$ is of the order of
isospin violation. In fact, eq.(\ref{viol}) tells us that 
even quite small values of these two
asymmetries could change significantly  the value of $s_W^2$
measured by NuTeV.

What do we know about the strange quark asymmetry?
An asymmetry $s^-$ of { the sign needed} to explain NuTeV 
can be induced non-perturbatively
({\it intrinsic strange}) by fluctuations of the kind
 $p\!\!\leftrightarrow\!\!\! \Lambda\, K^+$\cite{brodsky}. 
Unfortunately, the strange quark sea is
mainly constrained by (mostly old) $\nu N$ DIS data, which are
 usually not included in standard PDFs fits. In fact,  
MRST and CTEQ use an {\it ansatz}
$s\!=\!\bar{s}\!=\!(\bar{u}+\bar{d})/{4}$.
Barone et al. (BPZ) \cite{BPZ} have reanalyzed at NLO all $\nu N$ DIS 
together with  $\ell N$ and Drell-Yan data. They have 
a much higher sensitivity  to strange sea than the standard fits and 
find a strange $s(x)$ larger than usual at high-$x$. 
This feature contrasts  with NuTeV  dimuon results,
not included in the BPZ fit which was prior to their release, 
but agrees well with positivity constraints
from polarized DIS \cite{stefano}.
Allowing for a  strange asymmetry improves BPZ best fit drastically
and could explain a large fraction  
of the discrepancy. The result, 
 { $s^-\approx0.002$},
is compatible with theory estimates \cite{brodsky} and is driven by
cross section measurements by CDHSW ($\nu$N) and BCDMS ($\mu\,p$). 
\begin{figure}[t]
\center{\mbox{\epsfxsize=7.3cm\epsffile{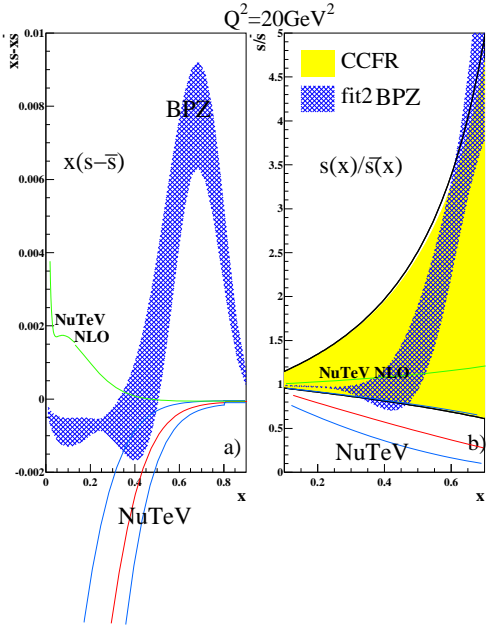}}}
\vspace{-2.1cm}
\caption{\sf Strange sea asymmetry at NLO from BPZ fit \cite{BPZ} (no
dimuons, blue band), at
LO in the NuTeV {\it cross section model}
from dimuons only \cite{nutev2} (red line, with errors in blue), 
and the same at NLO from \cite{nutev3} (green line, 
error not available). The yellow band on the rhs represents
the CCFR dimuon result.}
\vspace{-1cm}
\end{figure}

I have already mentioned that BPZ 
do not include NuTeV data, especially
those on dimuon events (tagged charm production),  a rather sensitive
probe of the strange sea.  NuTeV has analyzed them, claiming
$s^- =-0.0027\pm 0.0013$, which 
would {\it increase} the anomaly to 3.7$\sigma$ \cite{nutev2}. 
The NuTeV strange asymmetry is compared to the BPZ fit in Fig.~1, which
makes their incompatibility apparent.
Because of various shortcomings, such as strong dependence on
underlying PDFs, violation of strangeness (evident in Fig.~1) 
and other sources of
model dependence (see note added to \cite{noi}), the above estimate 
{\it cannot} be interpreted as a measurement of
$s^-$ and should not be compared to that of BPZ. 
NLO corrections, in particular, are very important for dimuons, as shown by  
a preliminary NLO analysis of NuTeV dimuons \cite{nutev3}. This new
analysis is in better agreement with BPZ, both on the total size 
and on the asymmetry of $s(x)$ (see Fig.~1). The use of the NuTeV  $s^-$
in the $s_\smallw^2$ extraction is also highly questionable, 
even in the context of NuTeV improved LO model, because 
it assumes that the restricted set of dimuon events
available be representative of the whole kinematic range employed in
the $s_\smallw^2$ analysis. At least, a generous theory
error should be attached to this procedure, perhaps of the order
of the effect itself, $0.7\sigma$, and   much  larger 
 than the theory error in eq.~(\ref{result}).

The bottom line is that 
we presently know very little on the strange sea. Before 
any conclusion can be drawn on  its asymmetry
and the effect on the NuTeV $s_W^2$   result, 
a global NLO fit including all dimuons and $\nu$-N DIS data is needed.
A precise $s(x), \bar{ s}(x)$ determination will be possible at a 
neutrino factory  \cite{nufact}.

A violation of isospin of the form $u_p(x)\neq d_n(x)$ would also 
affect the PW relation according to eq.~(\ref{viol}). A rough estimate
for its size is 
$(m_u-m_d)/{\Lambda_{QCD}}\approx  1\%$.
So small a violation of charge symmetry would give no  
visible effect in any  present experiment, apart from the NuTeV
measurement of $s_W^2$, where it could explain a fraction of the
anomaly -- about a third, according to eq.~(\ref{viol}).
Explicit model calculations \cite{isospin} vary widely in their
results for a shift in $s_W^2$. Estimates giving a very small shift
are generally due to subtle cancellations of much larger contributions
and should be handled with care. 

The relevant  momentum asymmetries in the quark sea 
are therefore only weakly constrained and could have a significant impact on
$s_W^2$ extracted by NuTeV. It  has been shown \cite{nutev2} 
that these effects are somewhat diluted in the actual
NuTeV analysis compared to the direct use of eq.~(\ref{viol}), 
precisely because NuTeV differs from a measurement of $R_{\smallp\smallw}$. 
They nevertheless introduce an unwelcome  uncertainty
very hard to estimate.

I should also mention that several attempts at explaining the NuTeV
anomaly with nuclear effects like nuclear shadowing
have been made \cite{nuclear}, but 
no convincing case has so far  been presented.

\section{ New Physics vs NuTeV}
A New Physics explanation of the NuTeV anomaly requires a $\sim 1$-2\% effect, 
and naturally calls for  tree level physics. It is very difficult to build
realistic models that satisfy all present experimental  constraints
and explain a large fraction of the anomaly \cite{noi}.

In particular, supersymmetry, with or without R parity, cannot help,
because it is strongly constrained by other precision
measurements (often at the permille level) 
and by direct searches.
The same is generally true of models inducing only oblique corrections
or only anomalous $Z$ couplings \cite{noi}. Realistic and well-motivated 
examples of the latter are 
models with $\nu_R$ mixing  \cite{noi,babu}.
Models with $\nu_R$ mixing {\it and} oblique corrections have been
considered in \cite{loinaz} and found to fit well all
data including NuTeV.
\footnote{Can the 
necessary oblique corrections be provided by a heavy SM Higgs boson?
No, as it is also clear from a careful reading of \cite{loinaz} 
(contrary to what stated by prominent NuTeV members,
there is {\it no} conflict between \cite{loinaz} and 
\cite{noi}). The only way  to obtain 
an acceptable fit with a preference for 
both $\nu$ mixing and  a heavy Higgs is to exclude $M_W$ from the data.
However,
solving the NuTeV anomaly at the expense of the very precise measurement of 
$M_W$ is hardly an improvement.} 
But finding sensible new physics that provides  
oblique corrections in the preferred range is far from obvious.

On the other hand, the required new physics can be parameterized  by
a contact interaction of the form
$ 
 [\bar{L}_2 \gamma_\mu L_2][\bar{Q}_1
\gamma_\mu Q_1]$. This operator might be induced by different kinds of
short-distance physics. Leptoquarks generally also induce another
operator which over-contributes to $\pi\to \mu \bar\nu_\mu$, or have
the wrong sign, but SU(2) triplet leptoquarks with non-degenerate
masses could fit NuTeV, albeit not very naturally.
Another possible new physics inducing
 the above contact interactions is
an unmixed $Z'$ boson. It could be either light ($2\lsim M_{Z'} \lsim
10$~GeV) and super-weakly coupled, or heavy ($M_{Z'} \gsim 600$~GeV). 
A viable possibility
that could alleviate  the NuTeV anomaly and 
at the same time explain  part of the $(g-2)_\mu$ anomaly \cite{anomuon}, is
based on an abelian gauge symmetries $B-3L_\mu$ \cite{noi}. 
The $Z'$ must have very small mixing with the $Z^0$ because of the
bounds on oblique parameters and on the anomalous $Z$ couplings 
\cite{noi,erler} (see  \cite{ma} for an explicit $L_\mu-L_\tau$ model and
\cite{chivu} for technicolor models).

\section{ The SM fit to $M_H$ is not satisfactory}
The global fit without NuTeV
 has an 11\% probability. This gives us an idea of the overall
 consistency of the data, but if we are interested in extracting
 information on the Higgs mass,  it is clear that we should
 concentrate only on the  subset of observables that 
are really { sensitive} to $M_H$ and, because of a strong correlation, 
to the top mass, $M_t$.
Using only $M_W$,$M_t$,$\Gamma_\ell$,
the $Z$-pole asymmetries, and   $R_b$, one obtains 
$\mh^{fit}=90$~GeV,  $\mh<195 $~GeV at 95\%~C.L.,
and  $\chi^2/$dof=13/4, corresponding to a  1\% probability.
In other words, the restricted fit gives the same constraints on $\mh$
 of the global fit. However, it is now obvious
 that the SM fit to the Higgs mass is {\it not}
satisfactory, even without NuTeV.

\section{ Another unwelcome anomaly}
The root of the problem is an  old $3\sigma$ discrepancy between 
the { Left-Right asymmetry, $A_{LR}$, measured by SLD
and  $A_{FB}^b$ measured by the  LEP experiments.
In the SM these asymmetries measure the {\it same} quantity, 
$\sin^2\theta_{eff}^{lept}$, related to the lepton couplings.
It now happens that all leptonic asymmetries, measured both at LEP and
SLD, are mutually consistent and prefer a very {\it light Higgs} mass. In
this sense, they are also consistent with $M_W$ measured at LEP and Tevatron.
Only the asymmetries into hadronic final states
prefer a {\it heavy Higgs} (see Fig.~2). 
\begin{figure}
\center{\mbox{\epsfxsize=6.6cm\epsffile{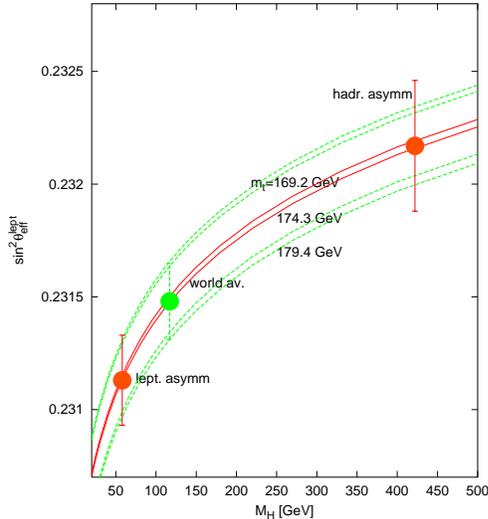}}}
\caption{\sf Higgs mass dependence of $\sin^2\theta_{eff}^{lept}$
extracted from leptonic and hadronic asymmetries,
for three $M_t$ values. 
}
\vspace{-.9cm}
\end{figure}

Since the hadronic asymmetries
are dominated by $A_{FB}^b$, and the third generation is naturally
singled out in many extensions of the SM, 
could this be  a signal of new physics in the  $b$ couplings?
After all, QCD and experimental systematics in $A_{FB}^b$ have been carefully
considered \cite{martin}.
New physics in the  $b$ couplings seems unlikely for several reasons:
(i) fixing  $\sin^2\theta_{eff}^{lept}$ 
at the value measured by the leptonic asymmetries, 
$A_{FB}^b$ corresponds to  a measurement of a combination of 
$b$ couplings, ${\cal{A}}_b=0.882\pm 0.017$; the same combination
is also tested by  $A_{LR}^{FB}$ at SLD, yielding  ${\cal{A}}_b=
0.922\pm 0.020$. 
One should compare these two values to the very precise SM prediction,
${\cal{A}}_b^{SM}=0.935\pm 0.002$: SLD result is compatible with the SM and at
1.5\,$\sigma$ from the value extracted from $A_{FB}^b$; 
(ii)  the value  of ${\cal{A}}_b$ extracted from $A_{FB}^b$
  would require a $\sim 30\%$ 
correction to the $b$ vertex, i.e.\  tree level physics;
(iii) $R_b$ agrees well with the SM and tests an orthogonal
combination of  $b$ couplings; it follows that  new physics should
predominantly affect the right-handed $b$ coupling, 
$|\delta g_R^b|\gg|\delta g_L^b|$. 
All this places strong restrictions on the extensions of the SM 
 that can explain 
$A_{FB}^b$.  Exotic scenarios that shift only the $b_R$ coupling include 
 mirror vector-like fermions  mixing with
$b$  quark \cite{wagner}, and 
LR models that single out the third generation \cite{valencia},
but even these {\it ad-hoc} models have problems in 
passing all experimental tests.
Difficult to explain in the most popular new physics models, both
NuTeV and $A_{FB}^b$ are in this sense two {\it unwelcome}
anomalies.

\section{ Too light a Higgs}
An even-handed option to handle the discrepancy
between $A_{LR}$ and $A_{FB}^b$
is to enlarge their error according to the  PDG prescription.
The result is a slight decrease in the central  $\mh$ value of the fit 
\cite{sirlin}. But we have seen that their preference for a heavy Higgs
really singles out the hadronic asymmetries.
It is then instructive to see what happens if one excludes
the hadronic asymmetries from the above restricted Higgs mass fit. 
Not surprisingly, a  consistent picture emerges: a very light Higgs with  
{ $\mh^{fit}=40$ GeV} fits perfectly all data and one obtains an upper
bound  $\mh<109 $~GeV} at 95\%~CL. 
If really $M_W$, $\Gamma_\ell$, $M_t$, and the leptonic asymmetries 
are consistent data and the SM is correct, 
why hasn't the Higgs been found at LEP, which set a lower bound
 $\mh> 114$~GeV \cite{chanowitz,altarelli}? 

The inconsistency with the direct lower bound marginally depends on
the value of the hadronic
contributions to $\alpha(M_Z)$ used in the fit, 
but even in the  most unfavorable
case the 95\%~CL upper bound is no more than 120~GeV. 
Similarly, current estimates of the theoretical error agree that 
it cannot shift up $\mh^{95\%}$ more than  $\sim 20$~GeV  
\cite{freitas}.
The inconsistency would be alleviated if the top mass turned
out to be heavier than the present central value, 
a possibility soon to be tested at Tevatron,
but the fit does not suggest this possibility at all. 
One can  quantify the inconsistency  computing the 
combined probability of the global fit {\it and} of having $\mh> 114$~GeV:
it is the same   with or without  $A_{FB}^b$
\cite{chanowitz}.

We have seen that excluding $A_{FB}^b$ and NuTeV from the  fit 
the quality of the fit
improves considerably, but $\mh^{fit}$ becomes very small.
Finding New Physics that simulates a very light Higgs is 
much easier than fixing the two anomalies.
An example are
oblique corrections: in general it just requires $S<0 (T>0)$ or 
$\epsilon_{2,3}<0$ \cite{chanowitz,altarelli}.
A non-degenerate unmixed fourth generation with a heavy neutrino with
$m_N\approx 50$~GeV would easily work \cite{novikov}.
More interestingly, the MSSM offers 
rapid decoupling (small corrections), 
  $\mw$ always higher than in SM, and 
 $\sin^2\theta_{eff}^{lept}$ lower than in SM.
A plausible MSSM scenario involves light sneutrinos and sleptons,
 heavy squarks, and $\tan\beta\gsim 5$ \cite{altarelli}.
The required mass spectrum 
 cannot be obtained in minimal SUGRA models with universal soft masses, though 
alternatives exist, and could be discovered at Tevatron.
Other susy scenarios have also been presented \cite{babu}.

\section{{Conclusions}}
The NuTeV experiment  aims at high precision 
in a { complex hadronic
  environment}. Its measurement of $\sin^2\theta_W$ is affected
by theoretical systematics  not fully  under control or untested, such as
a  small { strange/antistrange asymmetry}  and 
{ isospin violation}. The analysis should be upgraded to NLO.

Even excluding the NuTeV electroweak result, 
the SM fit to $M_H$ is not satisfactory. 
What we know on the Higgs boson mass  depends heavily on the $b$ quark 
FB asymmetries, an even more puzzling experimental anomaly.
Removing  the two deviant results from the SM fit leads however
to inconsistency with the direct lower bound on $M_H$. 

Both the NuTeV $\sin^2\theta_W$ and $A_{FB}^b$   require 
new {tree level}  effects which are difficult to accommodate in
reference scenarios of physics beyond the SM. 
 For instance, supersymmetry with or without R parity cannot explain them.
Proposed interpretations
rely on {\it ad-hoc} exotic models and
it is  always problematic to reconcile them with other precision data.
Keeping also in mind the discrepancy of the measured $(g-2)_\mu$ with the
SM prediction, the SM looks definitely under strain,  
although a clear-cut, compelling case for new physics has yet to be
made. 

\vspace{.5cm}

I am grateful to D.~Bardin for the invitation to ICHEP and to
S.~Davidson, S.~Forte, N.~Rius, and A.~Strumia
for a wonderful and stimulating collaboration and  
for many useful suggestions.

\end{document}